\documentclass[galaxies,article,accept,moreauthors,LaTeX and dvi2pdf]{Definitions/mdpi}
\firstpage{1} 
\pubvolume{xx}
\issuenum{1}
\articlenumber{5}
\pubyear{2020}
\copyrightyear{2020}
\history{Received: 20 July 2020; Accepted: 1 September 2020; Published: date}

\usepackage{subcaption}
\usepackage[countmax]{subfloat}
\Title {X-ray Flux and Spectral Variability of the TeV Blazars Mrk 421 and PKS 2155-304}

\Author{Alok C. Gupta} 

\AuthorNames{Alok C. Gupta}
\address [1]{%
Aryabhatta Research Institute of Observational Sciences (ARIES), Manora Peak, Nainital 263001, India; acgupta30@gmail.com} 

\corres{Correspondence: acgupta30@gmail.com}

\abstract{We reviewed X-ray flux and spectral variability properties studied to date by various X-ray satellites 
for Mrk 421 and PKS 2155-304, which are TeV emitting blazars. Mrk 421 and PKS 2155-304 are the most X-ray 
luminous blazars in the northern and southern hemispheres, respectively. Blazars show flux and spectral variabilities 
in the complete electromagnetic spectrum on diverse timescales ranging from a few minutes to hours, days, weeks, months 
and even several years. The flux and spectral variability on different timescales can be used to constrain the size of 
the emitting region, estimate the super massive black hole mass, find the dominant emission mechanism in the close 
vicinity of the super massive black hole, search for quasi-periodic oscillations in time series data and~several other
physical parameters of blazars. Flux and spectral variability is also a dominant tool to explain jet as well as
disk emission from blazars at different epochs of observations.} 


\begin{document}

\section{Introduction}

Blazar is a subclass of radio-loud (RL) active galactic nuclei (AGN) which includes BL Lacertae (BL Lacs) objects  
and flat spectrum radio quasars (FSRQs). Blazar’s central engine is a super massive black hole (SMBH) of the mass
range 10$^{6}$ M$_{\odot}$--10$^{10}$ M$_{\odot}$ that accretes matter and produces relativistic jets pointing 
almost in the direction of observer's line of sight \citep{1995PASP..107..803U}. Blazars show flux, spectral and 
polarization variability on all possible timescales ranging from a few minutes to several years across the entire 
electromagnetic (EM) spectrum \citep{1997ARA&A..35..445U}. The~emission from blazars in the complete EM spectrum 
is predominantly non-thermal. At~lower energies (from radio to soft X-rays), the~emission mechanism is certainly 
synchrotron radiation, while at high energies upto TeV energies the emission mechanism is probably due to inverse 
Compton (IC) emission e.g., \citep{2004NewAR..48..367K}, and~a hadronic origin is also plausible e.g., \citep
{2001APh....15..121M}.  

Blazars are among only a few classes of astronomical objects which emit radiation in the entire EM spectrum which 
gives an excellent opportunity to study their multi-wavelength (MW) properties, plot~the spectral energy distributions 
(SEDs) and~do the emission mechanism modeling. Blazar's SEDs from radio to $\gamma$-rays have two humps in the 
log($\nu F_{\nu}$) vs. log($\nu$) representation \citep{1997A&A...327...61G}. Based on the location of the first peak 
of SED, blazars are classified into low-energy peaked blazars (LBLs) and high-energy peaked blazars (HBLs) \citep{1995ApJ...444..567P}. 
In LBLs, the~first hump of SED peaks in the near-infrared (NIR)/optical, and~in the ultraviolet (UV) or X-rays for HBLs, while the second 
hump of SED usually peaks at GeV energies for LBLs and at TeV energies for HBLs. A~more recent classification of blazars
are suggested by \citep{2010ApJ...716...30A} based on the first SED peak frequency i.e.,~synchrotron peak frequency, 
$\nu_{peak}$, and~made the following classification: low synchrotron peak (LSP) blazars with $\nu_{peak} \leq$ 10$^{14}$ Hz; 
ISP~(intermediate synchrotron peak) blazars with 10$^{14} < \nu_{peak}$ 10$^{15}$ Hz; and  high synchrotron peak (HSP) blazars 
with $\nu_{peak} \geq$ 10$^{15}$ Hz.  

Variability in blazars on timescales of a few minutes to less than a day is often known as micro-variability 
\citep{1989Natur.337..627M}, intra-night variability \citep{1993MNRAS.262..963G} or intra-day variability (IDV) 
\citep{1995ARA&A..33..163W}; variability timescales from days to several weeks is called short timescale variability 
(STV), while variability on month to years timescales is known as long term variability (LTV) \citep{2004A&A...422..505G}.
Blazars are highly variable objects, so variability can be studied throughout different flux phases which can 
be considered to be: outburst, pre/post outburst and~the low-state.

TeV emitting blazars basically belong to HBLs or HSPs sub-classes. In~early 2000, only six TeV blazars (Mrk 421, 
Mrk 501, 1ES 1426$+$428, 1ES 1959$+$650, PKS 2155-304, 1ES 2344$+$514) were known; see \citep{2004A&A...422..505G} for a summary of their properties
. Thanks to the development of new $\gamma$-ray ground and space based facilities e.g.,~
{\it Fermi}\footnote{\url{https://fermi.gsfc.nasa.gov/.}}, High Energy Stereoscopic System {\it (HESS)} 
e.g., \citep{2003ICRC....5.2811H,2004APh....22..285F,2006A&A...457..899A}, Major Atmospheric Gamma-ray Imaging Cerenkov 
{\it (MAGIC)} e.g., \citep{2004NIMPA.518..188B,2005ICRC....5..359C,2006ApJ...638L.101A},~Very Energetic Radiation Imaging 
Telescope Array System {\it (VERITAS)} 
e.g., \citep{2006APh....25..391H,2008ICRC....2..747M,2008ICRC....3.1457M}, etc. These facilities have a very strong impact and made a 
complete revolution in TeV $\gamma$-ray astronomy. To~date, 70 TeV $\gamma$-ray emitting blazars have been discovered \footnote{\url{http://tevcat.uchicago.edu/.}}. A~few blazars including Mrk 421 and PKS 2155-304 were predicted to be potential 
neutrino emitting objects on the basis of protons accelerated in the cores can produce neutrinos if the soft radiation background in 
the core is sufficiently high \citep{2002PhRvD..66l3003N}. After~one and half decade of the prediction,~such a blazar TXS 0506$+$056 
is discovered \citep{2018Sci...361..147I}.   

In the present work, we discuss in detail the flux and spectral variability properties studied to date by various X-ray satellites for 
Mrk 421 and PKS 2155-304, which are TeV emitting HBL/HSP blazars. Mrk 421 and PKS 2155-304 are the most luminous X-ray blazars 
in the northern and southern hemispheres, respectively. 

The paper is organized as follows. In~Sections~\ref{sec2} and \ref{sec3}, we describe in detail the review of 
results obtained by various studies 
to date for the blazars Mrk 412 and PKS 2155-304, respectively. Section~\ref{sec4} provides the summary of this~work. 


\section{Review of Published~Results}
\label{sec2}
\vspace{-6pt}

\subsection{Mrk~421}
Mrk 421 (B2 1101+384; $\alpha_{2000.0} = \rm{11}^{h} \rm{04}^{m} \rm{27.2}^{s}$ and $\delta_{2000.0} = +\rm{38}^{\circ} \rm{12}{'} \rm{32}{''}$) at redshift \mbox{$z =$ 0.0308 \citep{1975ApJ...198..261U}} is one of the closest blazars with an intense point-like nucleus \citep{2008AIPC.1085..399W}.  Mrk 421 was discovered as an object with blue excess in the First Byurakan Survey which turned out to be an elliptical galaxy with a bright nuclear source classified as BL Lac object as it has a featureless optical spectrum, strongly polarized and displayed variable optical and radio fluxes and~has compact radio emission \citep{1972Afz.....8..155M,1975ApJ...198..261U,1975ApJ...201L.109M}. It~is located at a distance of 134 Mpc (H$_{0} = \rm{71} \ \rm{km} \ \rm{s}^{-1} \ \rm{Mpc}^{-1}$, $\Omega_{m} = \rm{0.27}$, $\Omega_{\lambda} = \rm{0.73}$). It is classified as a HBL/HSP blazar because the synchrotron peak of its SED is higher than 0.1 keV \citep{1992Natur.358..477P}.  By using various mass estimation methods, it is found that the mass of central SMBH of Mrk 421 is in the range of \mbox{(2--9) $\times$ 10$^{8} \ \rm{M}_{\odot}$} e.g., \citep{2002ApJ...569L..35F,2002A&A...389..742W,2003ASPC..290..621T,2008AIPC.1085..399W,2019ApJ...884..125Z}. It was detected in the GeV energy band by the Energetic Gamma Ray Experiment Telescope (EGRET) on board the Compton Gamma Ray Observatory (CGRO) \citep{1992ApJ...401L..61L,1992IAUC.5470....2M}.  It~is the the first known extra-galactic TeV $\gamma$-ray emitting object \citep{1992Natur.358..477P} and has been repeatedly confirmed as a TeV emitting source by various ground-based $\gamma$-ray telescopes e.g., \citep {1993AIPC..276..185S,2011ApJ...738...25A,2017ApJ...841..100A} and references therein. Mrk 421 is the brightest X-ray to $\gamma$-ray, emitting an extra-galactic object in the northern~hemisphere. 

\subsubsection{Flux and Spectral~Variability}

Mrk 421 was simultaneously observed for about 2 days in X-ray in the energy range 0.1--26 keV by {\it BeppoSAX} and TeV $\gamma$-ray at 
energies 0.5--2 TeV by {\it Whipple} in April 1998. A~large well correlated flare was observed in X-ray and $\gamma$-ray, which implied 
that X-ray and TeV photons derived from the same emitting region and from the same population of relativistic electrons \citep{1999ApJ...526L..81M}.
A coordinated X-ray observation of Mrk 421 was carried out from April 1997--May 1997 using {\it BeppoSAX} and {\it RXTE/ASM} covering the 
energy range 0.1--100 keV. The~source has shown significant flux and spectral variability on STV timescale and the spectrum in the energy 
range 0.1--100 keV has convex curvature which was interpreted in terms of synchrotron cooling \citep{2000MNRAS.312..123M}. In extensive
ten pointed observations of the blazar in 1997--1998 with {\it BeppoSAX} in the energy range 0.1--26 keV, the~following results were 
obtained: strong evidence of X-ray IDV, hard photons lag the soft ones by 2--3 ks, flare light curve is symmetric in the softest X-ray 
band but asymmetric at higher energies and~the peak of the synchrotron component shifted to higher energies during the rising phase 
and then receded \citep{2000ApJ...541..153F,2000ApJ...541..166F}. Flux and spectral variabilities of Mrk 421 using nine pointed observations 
taken from {\it XMM-Newton} from 25 May 2000 to 6 May 2004 were studied. On~all epochs of observations  the source has shown strong 
flux variability, and~different variability patterns, e.g., on one occasion the observed flux variation was more than a factor of three 
at highest X-ray energies and accompanied by complex spectral variations with only a small time lag between the hard and soft photons. 
The 0.2--10 keV spectra was well fitted by a broken power-law \citep{2001A&A...365L.162B}; 
spectrum were harder in a higher flux state, sometimes fitted by a broken power-law, and~sometimes very complex which could not be fitted 
by a broken power-law or a continuously curved model 
\citep{2002ApJ...574..634S,2003A&A...402..929B,2005A&A...443..397B}. By~using four pointed {\it XMM-Newton} observations of Mrk 421 in
November 2002, it was found that the source was highly variable, X-ray spectra were soft and steepened toward higher energies and~hardness 
ratio plots displayed a clear harder-when-brighter trend \citep{2004A&A...424..841R}. During~2003--2004 observations of the source
by {\it Rossi X-Ray Timing Explorer (RXTE)}, the~light curves show the presence of flares with varying amplitudes on a wide range of 
timescales, and~the X-ray spectrum becomes flatter \citep{2005ApJ...630..130B}. Mrk 421 data were taken over nine years (1997--2005) with {\it
ASCA, RXTE, EUVE, BeppoSAX and~XMM-Newton}. The~spectral evolution study of the source has shown that the SED has a lower peak at energies that 
vary in the range 0.1--10 keV while its X-ray spectrum is curved and fitted with a log-parabolic model \citep{2007A&A...466..521T}. In~March 
2001, a~week long coordinated observation in 2--60 keV X-rays {\it (RXTE)}, TeV $\gamma$-rays, and~optical showed strong variations in both 
X-ray and $\gamma$-ray bands, which are highly correlated with zero time lag. The~strong correlation further supports the standard model 
in which a unique electron population produces the X-rays by synchrotron radiation and the $\gamma$-ray component by IC scattering
\citep{2008ApJ...677..906F}. Simultaneous MW observations including X-rays in the energy range 0.3--200 keV by {\it Swift/XRT, 
INTEGRAL} and~{\it Swift/BAT}, radio, optical and~$\gamma$-rays were carried out for about 2 weeks in June 2006. Four strong flares at X-rays 
were observed that were not seen at other wavelengths (partially because of missing data). In~0.2--10 keV, data indicated a small correlation 
with the intensity which implied a hard-to-soft evolution but such correlation was missing in \mbox{20--150 keV}~\citep{2008A&A...486..721L}.          
In about a month long MW observing campaign in May--June 2008, in~hard X-rays (\mbox{20--60 keV}) {\it SuperAGILE} resolved a five-day
flaring event (9--15 June) which peaked at $\sim$55 mCrab. Data~from {\it SuperAGILE, RXTE/ASM} and {\it Swift/BAT} have shown a correlated 
flaring structure between soft and hard X-rays. A~{\it Swift/XRT} observation at flare detected the highest 2--10 keV flux ever observed from 
this source 2.6 $\times$ 10$^{-9}$ ers cm$^{-2}$ s$^{-1}$ \citep{2009ApJ...691L..13D}. In~X-ray observations taken during October 2005 to July
2006 in 0.2 to 50 keV, several episodic outbursts were seen \citep{2009ApJ...695..596H}. Spectral and flux evolution of Mrk 421 was studied
with {\it Swift} observations carried out during April to July 2006. In~this period, the~source exhibited both flux levels and SED peak energies 
each equal to their historic maximum until 2006. A~possible signature of acceleration processes that produce curved electron distributions was
found, and~the curvature decreases as the acceleration becomes more efficient \citep{2009A&A...501..879T}. In~a seven month long monitoring with 
the {\it MAXI GSC}, two strong X-ray flares from Mrk 421 were observed in 2--10 keV energy in January and February 2010. In~a February 2010
flare the observed flux was \mbox{164 $\pm$ 17 mCrab}, which was the highest among those reported from the object. A~comparison of the {\it MAXI}
and {\it Swift BAT} suggested a convex X-ray spectrum with photon index $\sim$$\Gamma \geq$ 2 \citep{2010PASJ...62L..55I}. A~MW
observing campaign of Mrk 421 from 2006 January to 2008 June was coordinated. Flux variability in the blazar was found in all EM bands 
except for the radio wave band. The~combing RXTE and Swift X-ray data show spectral hardening with increasing flux levels, and~in general 
correlated with an increase of the source activity in the TeV $\gamma$-rays \citep{2011ApJ...738...25A}. Coordinated observations with
{\it INTEGRAL}, {\it Fermi-LAT} and optical ground-based telescopes on 16--23 April 2013 were carried out. Two strong flares were detected  
in 3.5--60 keV from {\it INTEGRAL} and 0.1--100 GeV from {\it Fermi-LAT} observations, and~average flux in the 20--100 keV was 
$\sim$4.5 mCrab. The~time resolved spectra was fitted by broken power-law which was marginally better than the log-parabolic 
model \citep{2014A&A...570A..77P}. In coordinated observations of Mrk 421 between January 2009 and June 2009, a 
harder-when-brighter behavior in the X-ray spectra was found which also showed a strong correlation without any time lag between VHE $\gamma$-rays 
and X-ray fluxes \citep{2015A&A...576A.126A}. In~Mrk 421, a~flare occurred in March 2010 and was observed for 13 continuous days in the complete 
EM spectrum from VHE $\gamma$-rays to radio bands. A~remarkable flux variability was detected in X-ray and VHE $\gamma$-rays which slowly
decreased from high to low flux states \citep{2015A&A...578A..22A}. An~unprecedented double peaked outburst from 2013 April 10--16 was
detected with {\it NuSTAR} observation in 3--79 keV energies in which the first flare appears to have nearly a Gaussian shape with peak 
flux at $\sim$MJD 56395, while the second one, occurring two days later, was even  stronger with sharp rise and 
decay \citep{2017ApJ...841..123P}. {\it NuSTAR} observations were carried out in the historical low-flux state of Mrk 421 in 2013 January,
and for the first time a clear detection of a hard X-ray excess, above~$\geq$20 keV was found \citep{2016ApJ...827...55K}. A~MW
variability and correlated variability of Mrk 421 was carried out during exceptional X-ray flaring observed from 11--19 April 2013. Substantial flux variations on multi-hour and sub-hour timescales were observed in X-ray and $\gamma$-ray bands. Various X-ray and $\gamma$-ray 
bands were found to be well correlated without any time lag \citep{2020ApJS..248...29A}. A~detailed X-ray IDV study was carried out 
for Mrk 421 with 72 pointed observations from {\it Chandra} taken from 2000--2015 and 3 pointed observations from {\it Suzaku} taken during 
its whole operational period. Large amplitude IDV in soft and hard X-ray bands was detected. Variability time-scales ranging 
from 5.5 to 78.1 ks appeared to be present. Hard and soft bands were well correlated with zero time lag, and~in general harder-when-brighter 
trend in the spectral behavior was found \citep{2018MNRAS.480.4873A,2019ApJ...884..125Z}. { In~hardness ratio (HR) versus X-ray flux plots,
we noticed a clockwise as well as anti-clockwise loop at different epochs of observations which implied that particle acceleration as well as
synchrotron cooling both work in the source at different epochs of observations \citep{2018MNRAS.480.4873A}. In~a systematic study of the 16 year whole operation period of {\it RXTE}, 32 TeV blazar spectra were analyzed. From~photon spectral index ($\alpha$), flux, synchrotron radiation 
peak energy (E$_{p}$), electron spectral index ($p$) and~HR, it was found that when considering TeV blazars as a whole, $\alpha$ and X-ray 
luminosity are positively correlated, E$_{p}$ is negatively correlated with $p$ and $\alpha$ and~E$_{p}$ is positively correlated with HR
\citep{2018ApJ...867...68W}.}          

\subsubsection{Power Spectrum~Analysis}

For three pointed observations of Mrk 421 with {\it ASCA}, the power spectrum density (PSD) was plotted. The~best power-law slope ($\alpha$ 
value varies from 2.03 to 2.56, while on one occasion broken power-law was also fitted \citep{2001ApJ...560..659K}. Observation of the 
source with {\it MAXI} on another occasion was fitted with power-law slope $\alpha =$ 1.60, and~was also fitted with broken power-law 
\citep{2015ApJ...798...27I}. A~MW 
observing campaign of Mrk 421 was organized between January 2009 and June 2009, which included data from {\it VLBA, F-GAMMA, GASP-WEBT, Swift, RXTE, 
Fermi-LAT, MAGIC} and~{\it Whipple}. 
 PSD analysis on all wave bands were done and found that all PSDs can be described by power-laws without a break 
which is consistent with red noise behavior \citep{2015A&A...576A.126A}. In~3 pointed X-ray observations with {\it XMM-Newton} on 2014 April 29,  
 May 1--3, PSD distribution was at $\geq$4 $\times$ 10$^{-4}$ Hz and described by a power-law model with slope $\alpha =$ 1.2 to 1.8
\citep{2017ApJ...834....2A}. PSD analysis of long term {\it RXTE} and {\it Swift} X-ray observations of Mrk 421 were well fitted with power-law.
For {\it RXTE} and {\it Swift} X-ray light curves, the~PSD slopes were found to be 1.1 $\pm$ 1.6 and 1.3 $\pm$ 0.7, respectively \citep{2020MNRAS.494.3432G}.             
\subsubsection{Spectral Energy Distributions (SEDs)}

Simultaneous X-ray and $\gamma$-ray emission modeling of Mrk 421 revealed the first evidence for bulk jet Lorentz factors of the order of 
50 \citep{2001ApJ...559..187K}. About a decade long X-ray observations of Mrk 421 with {\it BeppoSAX, XMM-Newton} and~{\it Swift}
satellites in the energy range from 0.1 to over 100~keV were carried out. The~X-ray SED in different flux states was well fitted with a
log-parabolic model which also provided the good estimates of the energy and flux of the synchrotron peak in the SED. The~peak synchrotron
energy varies between 0.1--10 keV with different flux states \citep{2004A&A...413..489M,2008A&A...478..395M}. Multi-wavelength data taken 
at different flux states suggested that both SED peaks move to higher energies 
as the luminosity of the source increases; the measured SEDs failed to fit with one-zone synchrotron self-Compton (SSC) model, and~then by 
introducing an additional zone improves the fits \citep{2005ApJ...630..130B}. In MW data taken from radio to $\gamma$-ray 
bands in 2002 December to 2003 January, SED is fitted with an SSC model with very high Doppler factors and low magnetic fields 
\citep{2006ApJ...641..740R}. In~a week long MW campaign of Mrk 421 in March 2001, IDV on $\approx$30 min was found in
VHE ($>$200 GeV $\gamma$-rays), which was correlated with X-rays, but~not with the optical; the fractional variability increases from
optical to X-rays as a power-law. SED was well-fitted by the SSC model from cooling electrons injected with a Maxwellian distribution of 
characteristic energy \citep{2007A&A...462...29G}. Simultaneous MW SED during 2 weeks of observations in June 2006 were fitted
using a one-zone SSC model including the full Klein--Nishina cross section for IC scattering \citep{2008A&A...486..721L}. Simultaneous 
MW observations of Mrk 421 were carried out for two X-ray flarings in 2006 and 2008 in which SEDs were modeled using a leptonic 
model given by \citep{2002ApJ...581..127B}. It was found that a pure SSC model provides a good match to the SEDs during both observations
\citep{2009ApJ...703..169A}. A~4.5~month long multifrequency observational campaign was carried out for Mrk 421 in 2009 with {\it VLBA, 
Swift, RXTE, MAGIC}, the~{\it F-GAMMA, GASP-WEBT} and~other collaborations. During~this campaign, the~blazar showed a low flux activity 
at all wavelengths. The~MW SED was produced with a leptonic (one-zone SSC) and a hadronic model (synchrotron proton blazar)
\citep{2011ApJ...736..131A}. In the MW observing campaign of Mrk 421 from 2006 January to 2008 June, SED was generated for 18 
nights and well fitted by a one-zone SSC model \citep{2011ApJ...738...25A}. An intense MW monitoring of Mrk 421 was conducted from 
December 2007 until June 2008 with {\it MAGIC-I}, {\it Swift/XRT}, {\it Swift/UVOT} and other ground based data in radio and~optical 
bands. In the~obtained SED interpreted within the framework of a single-zone SSC leptonic model, a~high Doppler factor (40 $\leq \delta \leq$ 80)
was needed to reproduce the observed SED \citep{2012A&A...542A.100A}. A~detailed investigation of the electron energy distributions (EEDs) 
and the acceleration processes in the jet of Mrk 421 was carried out through fitting the SEDs in different flux states in the frame of the one 
zone SSC model. It was found that the shock acceleration is dominant in the low flux state, while stochastic turbulence acceleration is 
dominant in the flare state \citep{2013ApJ...765..122Y}. The~pre-flaring state of Mrk 421 on  22--23~March 2001 was observed in MW
and its SED was generated and fitted with so-called (lepto)hardronic models which are routinely used to model MW observations of 
HBL/HSP blazars. Using the ``leptohadronic pion" (LH$\pi$) model, the~X-rays are produced from the synchrotron radiation of a primary leptonic 
component while the $\gamma$-rays are pion induced. In~the ``leptohadronic synchrotron'' (LHs) model, the~X-rays are produced as from the synchrotron 
radiation while the $\gamma$-rays are produced by proton synchrotron radiation \citep{2013MNRAS.434.2684M}. In~a continuous 13 day observation
in March 2010, the one-zone SSC model can describe the SED of each day for the 13 consecutive days reasonably well while the flaring state was better 
described by a two-zone SSC model \citep{2015A&A...578A..22A}. In~an unprecedented double peaked outburst during 10--16 April 2013 observed
by {\it NuSTAR} in the energy range 3--79 keV, the~observed X-ray spectrum showed a clear curvature that was fitted by a log parabolic 
spectral form and is explained as originating from a log parabolic electron spectrum \citep{2015A&A...580A.100S}. Coordinated MW 
observations of Mrk 421 from radio to $\gamma$-ray energies during January--March 2013 were basically a quiescent state of the source
\citep{2016ApJ...819..156B} before an unprecedented double peaked X-ray outburst was observed by {\it NuSTAR} from 10--16 April 2013~\cite{2017ApJ...841..123P}. Both the synchrotron and IC peaks of the SED simultaneously shifted to frequencies below the typical quiescent state 
by an order of magnitude \citep{2016ApJ...819..156B}. Three days of coordinated X-ray and $\gamma$-ray observations on 2014 April 29, May~1--3
also included radio and optical archive data for broadband SED generation, and~the SED was found to be consistent with a one-zone SSC model 
\citep{2017ApJ...834....2A}.      

\section{PKS~2155-304}
\label{sec3}

PKS 2155-304 (H 2155-304; 
$\alpha_{2000.0} = \rm{21}^{h} \rm{58}^{m} \rm{52.07}^{s}$ and $\delta_{2000.0} = -\rm{30}^{\circ} \rm{13}' \rm{32.1}''$) was one of the 
first recognized BL Lac objects and was discovered as an X-ray source by the {\it HEAO 1} X-ray 
satellite \citep{1979ApJ...229L..53S,1979ApJ...234..810G,1980ApJS...43...57H}. It is like most other BL Lac objects associated with a compact, 
flat spectrum radio source, and~has an almost featureless continuum from radio to X-ray energies. It is the most luminous object from UV to TeV 
$\gamma$-ray energies in the southern hemisphere. The~redshift of PKS 2155-304 was estimated to be\mbox{ $z =$ 0.116 $\pm$ 0.002} by optical 
spectroscopy of the galaxies in the field of the BL Lac object \citep{1993ApJ...411L..63F}. {\it EGRET} on board the {\it CGRO} detected 
$\gamma$-ray emissions from the source in the energy range from 30 MeV to \mbox{10~GeV \citep{1995ApJ...454L..93V}}. It~was detected in TeV $\gamma$-ray
energies by observations from Durham Mark 6 Telescopes \citep{1999ApJ...513..161C}.   

\subsection{Flux and Spectral~Variability}

{\it EXOSAT} observed PKS 2155-304 in X-ray energies at nine epochs in 1983--1985. Quasi-simultaneous observations of the source were also
carried out in far-UV with {\it IUE} and optical/NIR with ESO telescopes. On~two occasions the rapid flux rising was observed with a doubling
time $\sim$1~h, and~the X-ray spectra in the energy range (1--10 keV) were well fitted with single power-law plus 
absorption \citep{1989ApJ...341..733T}. Detailed hard X-ray properties of PKS 2155-304 based on observations were made in 1988 and 1999 with 
the {\it Large Area Counter (LAC)} on board the {\it Ginga} satellite. The~source exhibited large variability of a factor of 7 in the energy
range 2--6 keV. The~intensity decline of a factor of 2 in amplitude within 4 h in this energy range. The~X-ray spectrum characterized by
a break at $\sim$4~keV and it hardens as the intensity increases \citep{1993ApJ...404..112S}. Extensive {\it ROSAT PSPC} observations of 
the source taken during 12--15 November 1991 revealed that it was in a bright flux state, and~rapid X-ray flux variation upto 30\% in a day
was detected. The~soft X-ray flux was correlated with simultaneous UV flux taken with {\it IUE}, and~the soft X-ray spectrum remained 
unchanged during the whole duration of observations. Individual {\it ROSAT PSPC} spectra was well fitted with single power-law with photon
index \mbox{$\Gamma \sim -$2.65 \citep{1994A&A...288..433B}}. A~simultaneous MW observing campaign of PKS 2155-304 was carried out in November 
1991 in X-ray, UV, optical, IR and radio bands. Fluxes in X-ray, UV and~optical bands were strongly correlated, with~the X-ray leading UV,
optical by 2--3 h. UV and Optical fluxes showed variation of a factor of $\sim$2 in a week time, while X-ray/UV/optical showed 
$\sim$10\% changes in flux in a few hours \citep{1995ApJ...438..120E}. In~a 10~day  simultaneous MW campaign of the blazar in May 1994,
the source was observed in X-rays by {\it ASCA} and {\it ROSAT} X-ray satellites. The~X-ray light curve showed a well-defined X-ray flare.
The~X-ray flare observed with {\it ASCA} showed a factor of 2 flux increase in about half a day and decayed roughly as 
fast \citep{1997ApJ...486..799U}. In 100 ks observations with {\it BeppoSAX} in the energy range 0.1--100 keV, the~source was detected in an 
intermediate intensity level compared to previous observations. A~number of spectral features detected with observation which was well 
described by a convex spectrum with (energy), and~slope gradually steepening from 1.1 to 1.6 \citep{1998A&A...333L...5G}. A~pointed
observation of the source from {\it BeppoSAX} was carried out continuously for about 1.5 days beginning on 22 November 1997. The~light
curves indicated that the X-ray flux was close to the highest detected level and higher by a factor of 2 than that observed by {\it BeppoSAX} 
in 1996. The~X-ray spectra showed a curved continuum, with~no evidence of spectral features, extended up to $\sim$50 keV 
\citep{1997IAUC.6776....2C,1998A&A...333L...5G,1999ApJ...521..552C}. Four~pointed observations of PKS 2155-304 were carried out during
1994--1999 with {\it ASCA} and {\it BeppoSAX}. On~a timescale of less than an hour, no large amplitude-variability event was detected, 
the light curves in different X-ray energy bands were found to be highly correlated without any time lag and the amplitude of variability 
increased with energy \citep{1999ApJ...527..719Z,2002ApJ...572..762Z}. Time-resolved spectra fitted with a curved model suggested that the 
peak position of synchrotron emission shift to higher energy with increasing flux, spectral changes are complicated and there were no clear
correlations of spectral slope versus flux and between spectral slopes at different energies \citep{2002ApJ...572..762Z}. Extensive X-ray 
studies of PKS 2155-304 with {\it XMM-Newton} satellite data were carried out in a series of 
papers \citep{2005ApJ...629..686Z,2006ApJ...637..699Z,2006ApJ...651..782Z}. Extensive study of {\it XMM-Newton}
provided the following results: (i) the excess variance (absolute rms variability amplitude) and the fractional rms variability amplitude show 
linear correlation with source flux, (ii) using the normalized excess variance, the~black hole mass of PKS 2155-304 was estimated to
be 1.45 $\times$ 10$^{8}$ M$_{\odot}$, (iii) the hardness ratio versus flux plots showed that the spectral changes were mainly significant
during flares, (iv) the cross-correlation of the light curves in different energies were well correlated with different time lags, (v) the
source has shown large amplitude X-ray IDV \citep{2005ApJ...629..686Z,2006ApJ...637..699Z,2006ApJ...651..782Z}. By~using 20 {\it XMM-Newton} 
archival observations of PKS 2155-304 taken from 2000 to 2012, long term flux and multi-band cross-correlated variabilities were studied.
Significant flux variations were observed in all optical, UV and X-ray energies. Optical and UV bands data were well correlated while soft
and hard X-ray energies light curves were well correlated which suggests that the optical/UV and X-ray emissions in this source may arise 
from different lepton populations \citep{2014MNRAS.444.3647B}. There were three continuous pointed observations of PKS 2155-304 on 
24 May 2002 with {\it XMM-Newton}. These observations display a mini–flare, a~nearly constant flux period, a~strong flux increase. A~time
resolved cross-correlation analysis between different X-ray bands detected significant hard and soft lags (for the first time in a
single observation of this source) \citep{2016NewA...44...21B}.                 

\subsubsection{Spectral Energy Distributions (SEDs)}

A simultaneous MW observing campaign of PKS 2155-304 was carried out in November 1991, and~SED was tried to fit with various standard 
models e.g.,~the synchrotron/Compton models, accretion disk model and~gravitational lensing model. None of these model could satisfactorily
explain the findings \citep{1995ApJ...438..120E}. The~broad band SED generated with the MW campaign of the blazar in May 1994 was fitted 
with various models. The~SED temporal profile fitted with the synchrotron emission from an inhomogeneous, relativistic 
jet \citep{1997ApJ...486..799U}. A~simultaneous MW observation on 22 November 1997 broad band SED was well fitted with a one-zone SSC
model \citep{1999ApJ...521..552C}. Using two X-ray pointed observations of the source with {\it XMM-Newton}, the~first evidence of IC 
X-ray emission below 10 keV from the source was found, spectra in 0.6--10 keV harden ($\Delta \Gamma \sim$ 0.1--3) at break energy 
$\sim$4 keV, and~the quasi-simultaneous optical/UV/X-ray SEDs suggested concave X-ray spectra of the source \citep{2008ApJ...682..789Z}. 
In an MW campaign from the radio to X-ray bands in 1994 May, a~time-dependent SSC model for flare provided B $\sim$0.1--0.2 G, and~
relativistic beaming with a Doppler factor of $\delta \sim$ 20--30 \citep{2000ApJ...528..243K}. MW observations of PKS 2155-304
by the {\it Swift} satellite and other EM band data from ground-based telescopes at the end of 2006 July reported the dramatic 
increase in TeV flux; the~X-ray flux changed by a factor 5 without a large spectral change. SED modeling based on the SSC process in a 
homogeneous region suggested the Doppler factor $\delta =$ 33 \citep{2007ApJ...657L..81F}. An MW observing campaign of PKS 2155-304 was conducted 
from 25 August 2008 to \mbox{6 September} 2008 with {\it Fermi, HESS, RXTE, Swift} and~ATOM. Contrary to previous findings in flaring state,
no strong correlation was found in X-ray and VHE $\gamma$-rays, although~the SSC model nicely fitted MW SED \citep{2009ApJ...696L.150A}.
The two week long MW observations of the blazar in July and August 2006 was the~period when two exceptional VHE $\gamma$-ray flares occurred.
X-ray and VHE $\gamma$-ray emission were found to be correlated during the observed flaring state of the source. The~nightly averaged high-energy 
spectra of the non-flaring nights were reproduced by a stationary one-zone SSC model, with~only small variations in the parameters. The~
spectral and flux evolution in the high-energy band during the night of the second VHE flare were modeled with the multi-zone SSC model 
\citep{2012A&A...539A.149H}. By~using 20 {\it XMM-Newton} archival observations of PKS 2155-304 taken from 2000 to 2012, simultaneous
optical, UV and X-ray SEDs were generated for individual observation. The~SEDs were fitted well with the power-law + log-parabola (PLLP) model
\citep{2014MNRAS.444.3647B}. X-ray emission of PKS 2155-304 during different flux states in 2009--2014 were studied with {\it XMM-Newton}
archive data. Spectral curvature of most of the observations showed curvature or deviation from a single power-law and can be well modeled 
by a log parabola model \citep{2017ApJ...850..209G}. 

\subsubsection{Power Spectrum~Analysis}

In three {\it EXOSAT} observations of PKS 2155-304 when it was in a flaring state, 6--7 November 1984 and 24 October 1985, PSDs were calculated
and an average power-law slope of about $-$2.5 was obtained for the energy range 1--6 keV \citep{1991ApJ...380...78T}. A~detailed PSD analysis of 
the X-ray light curves of PKS 2155-304, observed with {\it BeppoSAX} and {\it ASCA} in 1994 to 1999 was carried out. From~{\it ASCA}, pointed 
observations of the source were carried out for continuous $\sim$50 h on 19--21 May 1994, while {\it BeppoSAX} pointed observations were carried 
out for $\sim$60 h on 20--22 November 1996, $\sim$35 h on 22--24 November 1997 and~$\sim$62 h on 4--6 November 1999. PSDs were 
fitted with power-law and their slopes were found to be in the range of 1.54 $\pm$ 0.07 to 3.10 $\pm$ 0.76 \citep{1999ApJ...527..719Z,2002ApJ...572..762Z}.    
A detailed PSD analysis was carried out for fifteen pointed observations of PKS 2155-304 with {\it XMM-Newton} in the energy range 
(0.3--10~keV) taken from 2000 to 2008. PSDs fitted with power-law and with a large range of the slope from \mbox{$-$3.52 $\pm$ 0.76} to \mbox{$-$1.10 $\pm$ 0.48} were 
estimated \citep{2009A&A...506L..17L,2010ApJ...718..279G}. On~another occasion, PSD analyses of eleven pointed observations of PKS 2155-304 
with {\it XMM-Newton} were conducted in the energy range 0.2--10 keV since its launch to 2011. PSDs were well fitted with power-law, and~their
slopes were found in the range of 2.1--2.3~\citep{2012A&A...544A..80G}. 
PSD~analysis of long term {\it Swift} X-ray observations of PKS 2155-304 fitted well with 
power-law, and~its slope was found to be 1.3 $\pm$ 2.1 \citep{2020MNRAS.494.3432G}.

\subsubsection{Quasi Periodic Oscillation (QPO)}

Detection of periodic and/or quasi periodic oscillation (QPO) in the light curve of blazars is very rare and occasional e.g., see for review \citep{2014JApA...35..307G,2018Galax...6....1G}, and references therein. PKS 2155-304 is one of a few blazars which have shown evidence of QPO
detection on diverse timescales in some EM bands. {\it International Ultraviolet Explorer (IUE)} observations of PKS 2155-304 in UV band
have shown a short lived QPO of period $\sim$0.7 day \citep{1993ApJ...411..614U}. Using $\sim$17 years of miscellaneous data in optical
UBVRI bands, evidence of QPO detection with a period of 4 and 7 years was found \citep{2000A&A...355..880F}. A~strong evidence of $\sim$4.6 h 
QPO was found in the source in {\it XMM-Newton} observations made on 1 May 2006 \citep{2009A&A...506L..17L}, and~on another occasion a weak QPO 
in the source with period 5.5 $\pm$ 1.3 ks was reported in a {\it XMM-Newton} observation made on 24 May 2002 \citep{2010ApJ...718..279G}. Using
the X-ray QPO period of 4.6 h, the~super massive black hole mass for PKS 2155-304 was found to be 3.29 $\times$ 10$^{7}$ M$_{\odot}$ for
a non-rotating (Schwarschild) black hole and \mbox{2.09 $\times$ 10$^{8}$ M$_{\odot}$} for a maximally rotating (Kerr) black hole \citep{2009A&A...506L..17L}. 
Using long term optical/NIR inhomogeneous data collected from various archive and published literature, QPO was detected of $\sim$315 $\pm$ 5 days
on a few occasions \citep{2014RAA....14..933Z,2014ApJ...793L...1S,2016AJ....151...54S}. Using {\it Fermi-LAT} observations in 100 MeV--300 GeV
taken from August 2008 to May 2014, a~$\gamma$-ray QPO with period of $\sim$642 days was estimated \citep{2016AJ....151...54S}. Using~a longer
data train of {\it Fermi-LAT} taken from August 2008 to 2016 October, $\gamma$-ray QPO with a period of \mbox{1.74 $\pm$ 0.13} years was detected 
\citep{2017ApJ...835..260Z} which confirmed the QPO detection period of \citep{2016AJ....151...54S}. This is possibly the only evidence of
QPO detection in optical polarization with the period of 13 min in the source which is the only optical polarized QPO detection in any 
AGN to date \citep{2016MNRAS.462L..80P}.    


\section{Summary}
\label{sec4}

Flux and spectral variability of blazars is one of the most frequently used tools to understand the emission mechanisms responsible in 
different EM bands at different flux states, to~determine size of emitting, estimating central super massive black hole mass, cross
correlated flux variability, spectral variation and~the geometry of the local regions. {In the present work we have extensively
searched for the important results mainly based on X-ray flux and spectral variabilities of the blazars Mrk 421 and PKS 2155-304.
The results of both of these blazars are summarized~below.   

\begin{itemize}
\item[$\bullet$] X-ray flux variability of both of these blazars is very complex, with~patterns changing from epoch to epoch, and~also depending 
on their flux~states. 
\item[$\bullet$] In general, the X-ray spectra of both of these blazars is well fitted by log-parabolic~model. 
\item[$\bullet$] In both the blazars, X-ray spectra in general harden with increasing flux but occasionally the opposite trend is also~found. 
\item[$\bullet$] In both the blazars, X-ray spectra has shown convex curvature which was interpreted in terms of synchrotron~cooling. 
\item[$\bullet$] Synchrotron SED hump of both of these blazars peaking from 0.1 to 10 keV depending on their flux state of the~source. 
\item[$\bullet$] PSDs of X-ray light curves of both of these blazars are red noise dominated and in general well fitted with power-law, but occasionally broken power-law can give a better~fit.   
\item[$\bullet$] On some occasions, for both the blazars, well correlated flare in X-ray and $\gamma$-ray are found which implied that X-ray and $\gamma$-ray photons derived from the same emitting region and from the same population of relativistic electrons. Such correlation further supports the standard model in which a unique electron population produces the X-rays by synchrotron radiation and the $\gamma$-ray component by IC~scattering. 
\item[$\bullet$] For both the blazars, even in  general for TeV emitting blazars, hardness ratio (HR) versus X-ray flux shows a clockwise and anti-clockwise loop which implies particle acceleration as well as synchrotron cooling work at different epochs of~observation. 
\item[$\bullet$] For both the blazars, even in  general for TeV emitting blazars, it is found that photon spectral index ($\alpha$) and X-ray luminosity are positively correlated, synchrotron radiation peak energy (E$_{p}$) is negatively correlated with electron spectral index ($p$) and $\alpha$ and~E$_{p}$ is positively correlated with~HR.
\item[$\bullet$] For high and rapid $\gamma$-ray flux variability, an~extremely high value of Doppler factor (40 $\leq \delta \leq$ 80) was needed for Mrk~421. 
\item[$\bullet$] For high and rapid $\gamma$-ray flux variability, a~high value of Doppler factor upto 30 was~needed.
\item[$\bullet$] In general, the MW SEDs of both of these blazars are well fitted with the one-zone SSC~model.
\item[$\bullet$] In unprecedented X-ray flare detection in both these blazars, the~SEDs are better fitted with two-zone SSC~model.
\item[$\bullet$] On some peculiar variable nature of light curves, the~combination of SSC, EC and IC models may better explain MW SEDs of both of  these~blazars. 
\item[$\bullet$] When Mrk 421 shows unprecedented strong $\gamma$-ray emission, then the MW SED is better explained by leptohadronic 
pion and/or the leptohadronic synchrotron~model.
\item[$\bullet$] Hard X-ray excess above $\geq$ 20 keV was detected in Mrk~421.
\item[$\bullet$] PKS 2155-304 is one of a few blazars which has shown QPOs on diverse timescales in different EM bands.
\end{itemize}
\vspace{12pt} 

\funding{This research received no external funding.}

\acknowledgments{ I thankfully acknowledge the reviewers for very useful comments which helped to improve the~manuscript.}
%
\conflictsofinterest{The author declares no conflict of~interest.} 



\reftitle{References}


\end{document}